**Guns, Zombies, and Steelhead Axes: cost-effective recommendations for surviving human societies**

Authors: Jacob Pacheco[1#], Ahmani Roman[2#], Kursad Tosun[1], Scott Greenhalgh[1*]

1. Department of Mathematics, Siena College, 515 Loudon Road, Loudonville, NY, 12211
2. Department of Physics, Siena College, 515 Loudon Road, Loudonville, NY, 12211

*-corresponding author
#-authors contributed equally

Word count: 5563

## Abstract

In pop culture, there are many strategies to curtail zombie apocalypses. However, it remains unclear what routes we should take to eliminate zombies effectively and affordably. So, we created a mathematical model to examine interventions that armed adults with steelhead axes or provided enough ammunition and a 9mm handgun to kill zombies in either a "single-tap" or "double-tap". We investigate each case over two years under slow, moderate speed, and fast zombies scenarios. We quantify health burden by zombies averted, disability-adjusted life-years averted, and determine cost-effectiveness using the incremental cost-effectiveness ratio. Our predictions show the single-tap intervention is the best for stopping the zombie apocalypse, as it would avert hundreds of millions of zombies and deaths while also being the most cost-effective intervention. Altogether, this suggests conserving ammunition and supplying ranged weapons would be an effective use of limited resources in the event of a zombie uprising.

**Introduction.**

Zombies! An undying fascination of humans worldwide for decades, and a theoretical disease of which the world has never encountered. Simply put, eliminating zombies would be a daunting, if not impossible task, at least according to pop cultures, such as *The Walking Dead*, *Zombieland*, *World War Z*, and others [1–3]. Attempts to eliminate zombies in pop culture have included the deployment of the military, the use of nuclear bombs, and vaccination of the entire population with non-fatal diseases [4,5], all of which have had varying degrees of success.

While each pop culture intervention has its merit concerning eliminating zombies, they also pose various degrees of risk to public health. For instance, the use of nuclear weapons would almost surely have negative health consequences for any surviving humans near their deployment, and vaccination with pathogens could have detrimental effects on its long-term evolution [6,7]. Fortunately, mathematical models are commonplace for predicting disease outbreaks [8,9], with numerous works dealing with predicting the trajectory of a zombie apocalypse [10–12]. Such models are also commonly used to measure the number of avoidable deaths and the average years of life lost because of the health burden of infectious diseases. However, while mathematical models exist [10–12] to predict the potential effect of a zombie outbreak globally, their use to inform on effective and affordable interventions to inhibit the spread of zombies remains unaddressed. Therefore, to inform this issue, we developed a zombie outbreak model to evaluate interventions for controlling zombie outbreaks and quantify their health benefits and costs for society.

As seen in pop culture, the speed of zombies varies substantially and plays a crucial role in the level of danger they pose. We consider zombies moving at slow, moderate, and fast speeds and develop a mathematical model to estimate the health benefits and costs of interventions that arm adults with steelhead axes, 9mm handguns with the ability to kill zombies with one shot, or two shots, respectively. Specifically, we measure the overall burden of zombies by disability adjusted-life-years (DALYs) [13], determine the cost-effectiveness of each intervention using the incremental cost-effectiveness ratio (ICER) [14], in addition to estimating the zombies and deaths averted for each intervention.

In zombie media, it is common to see melee weapons such as swords, knives, and axes used to fight zombies. We discovered that for the general adult population, melee weapons are unsafe, it is safer for adults to avoid close-quarter combat with zombies. Instead, our mathematical model predicted that an intervention where 9mm handguns are administered to the population with the ability to kill zombies with one bullet would be ideal. We determined which intervention would be the most beneficial by quantifying the number of zombie cases, human deaths, and DALYs averted annually following the administration of weapons into the adult population.

**Methods**.

To evaluate the health benefits and cost-effectiveness of interventions aimed towards eliminating zombies in the event of an apocalypse, we created a mathematical model calibrated to the current population demographics of the United States. We use the model to estimate the effects of interventions on the number of zombies and deceased individuals. We calibrate our model on data provided by zombie apocalypses in pop culture and other comparable diseases and consider three interventions scenarios over two years. Specifically, we explore interventions of arming the adult population with steelhead axes and arming the adult population with 9mm handguns using either the "single-tap" (ST) or "double-tap" (DT) [15]. intervention for eliminating zombies. In addition, we predict the DALYs and ICER associated with further information on the health benefits and costs of each intervention.

**2.1 SZD model.** We developed a compartmental model to predict the spread of a zombie outbreak throughout the United States. The model considers three main compartments; susceptible individuals, $S$, zombies, $Z$, and deceased individuals, $D$. The S compartment is also subdivided by age into three categories, children, $C$, adults, $A$, and the elderly, $E$. The rates of transition between these compartments are given by the following system of differential equations:

$$\frac{dC}{dt} = \omega_3 E - \frac{\beta}{H} CZ - \omega_1 C,$$

$$\frac{dA}{dt} = -\frac{\beta}{H} AZ + \omega_1 C - \omega_2 A - \frac{\beta_{weapon}}{H} AZ,$$

$$\frac{dE}{dt} = -\frac{\beta}{H} EZ + \omega_2 A - \omega_3 E,$$

$$\frac{dZ}{dt} = \frac{\beta}{H}(C + A + E)Z - \alpha_{baseline} Z + \frac{\beta_{weapon}}{H} AZ - \frac{\alpha_{weapon}}{H} AZ$$

$$\frac{dD}{dt} = \alpha_{baseline} Z + \omega_3 E + \frac{\alpha_{weapon}}{H} Z,$$

here, $b$ is the birth rate, $\beta$ represents the transmission rate of the zombieism, $\beta_{weapon}$ is the increase in transmission for humans that either use the ST, DT, or an ax, H is the total population of humanoids (humans and zombies), $\alpha_N$ is the natural rate that humans kill zombies, and $\alpha_{weapon}$ is the rate humans kill zombies using either an ST, DT, or an ax. Note, $\beta_{weapon}$ and $\alpha_{weapon}$ are both considered to be zero in the baseline. In addition, the rates of the transitions between age classes in the susceptible compartment are governed by the parameters $\omega_i$, for i = 1, 2, 3.

**2.2 Parameter estimation.** For our model, we assume zombies' have a natural death rate because studies have shown that zombies die without food after 20 days [16]. Thus, we take the natural death rate of zombies to be $\alpha_{baseline} = \frac{1}{20} \ days^{-1}$. To compute the transmission rate $\beta$, we made use of its connection to the basic reproductive number of a disease, as determined by the next generation approach. Specifically, for our model, it follows at the zombie-free equilibrium (ZFE) that

$$C_0 = \frac{w_3 w_2 N}{(w_3 w_1 + w_3 w_2 + w_2 w_1)},$$

$$A_0 = \frac{w_3 w_1 N}{(w_3 w_1 + w_3 w_2 + w_2 w_1)},$$

$$E_0 = \frac{w_1 w_2 N}{(w_3 w_1 + w_3 w_2 + w_2 w_1)}.$$

Thus, we have that

$$F|_{ZFE} = \beta,$$
$$V|_{ZFE} = \alpha_{baseline}.$$

It follows that the basic reproductive number is

$$R_0 = FV^{-1} = \frac{\beta}{\alpha_{baseline}}.$$

Thus, we obtain

$$\beta = \alpha_{baseline} R_0.$$

Finally, to determine $\beta$, we considered $R_0$ values that describe scenarios of slow, moderate, or fast spread of zombies based on the highly communicable diseases of rubella, $R_0 = 7$, chickenpox, $R_0 = 12$, and measles, $R_0 = 15$ [17,18].

**Table 1. Parameters, values, and sources.**

| Definition | Parameter | Value | Citation |
|---|---|---|---|
| The population of the United States | $N$ | 332807303 | [19] |
| Average duration of childhood | $1/\omega_1$ | 18 years | [19] |
| Average duration of adulthood | $1/\omega_2$ | 42 years | [19] |
| Average duration of being elderly | $1/\omega_3$ | 19 years | [19] |
| Contact rate | $\sigma$ | 15.18 zombies per day | [20] |
| The basic reproductive number of zombieism | $R_0$ | 7, 12, 15 | [17,18] |
| The natural rate zombies die | $\alpha_{baseline}$ | 0.05 per day | [21] |
| The rate adults kill zombies during the ax intervention | $\alpha_{ax}$ | 0.101 per day | Sec 2.3 |
| The rate adults kill zombies during the ST intervention | $\alpha_{ST}$ | 6.101 per day | Sec 2.3 |
| The rate of adults kills zombies during the DT intervention | $\alpha_{DT}$ | 10. per day | Sec 2.3 |
| The transmission rate of zombieism | $\beta$ | $\alpha_{baseline} R_0$ | Sec 2.1 |
| DALY discount rate | $r$ | 0.03 | [22] |
| Disability weight from being a zombie | $D_z$ | 0.471 | [13] |

**2.3 Zombie interventions.** For our baseline, we assume that humans are typically ill-equipped and unable to effectively kill zombies. We evaluate the effects of three interventions that improve humans' ability to kill zombies during a zombie apocalypse. Specifically, supplying adults with steelhead axes, a 9mm handgun with ammunition and recommending the ST intervention, and finally providing adults with a 9mm handgun with ammunition and recommending the DT intervention. We explored the effect of each of these interventions over two years on their ability to mitigate zombie apocalypses that feature slow, moderate, and fast spread of zombieism as measured by the number of zombies, deaths, and DALYs averted. We base our weighting of zombification on neurological sequelae, life-long cognitive issues stemming from prior disease[22].

**2.3.1. The steelhead ax intervention.** For the steelhead ax intervention, we assume only adults are provided with a steelhead ax. Furthermore, based on video evidence[16], we assume the average adult has the upper limit of killing 67 zombies in 2 minutes before physical exhaustion. Combining this fact with the average American physical exercise of 20 min three days a week, we get that the average person should be capable of killing zombies 8.5 mins per day, which yields a maximum of 287.14 zombies killed per day with an ax. On average, an adult encounters other humans and zombies at the contact rate $c = 30.36$ contacts per day [20], and so when armed with an ax, they can kill a zombie at a rate of $\alpha_{ax} = \frac{30.36}{287.14} = 0.106$ corpses per day. In addition, due to the close-quarter nature of using an ax to kill zombies, we also assume that adults are placed at a heightened risk for transmission,

$$\beta_{ax} = \beta_{baseline} \cdot p_{ax\ accident},$$

To determine the probability of an ax accident $p_{ax\ accident}$, we used recent data on workplace injuries. Specifically, we assumed killing each zombie with an ax would present the same level of risk as workplace injuries. As such, the Bureau of Labor statistics workplace injury ratio in 2018 was 3.5 cases per 100 employees. Hence, the odds of an accident in killing a maximum of 287.14 zombies in 1 day is $0.035$. So, the increase in transmission rate is $\beta_{ax} = \beta_{baseline} \cdot 0.035$.

**2.3.2. The single-tap intervention.** The ST intervention arms civilian adults with a 9mm pistol, under the premise of using one gunshot to kill one zombie. Implicit in this setup, is the assumption that military personnel and police officers, which comprise $p_{police} = 0.01$ of the adult population[23], are already armed.

We assume that the contact rate for humans and zombies combined is $c = 30.36$ contacts per day[20] therefore, we take the zombie contact rate as $30.36 \cdot \frac{Z}{H}$[20]. Thus, in order to have enough ammunition to deal with zombie encounters, adults need approximately 31 bullets per day.

To estimate the rate zombies are killed under the ST intervention $\alpha_{ST}$, we used a study on police officer shooting accuracy in New York City and Los Angeles [24]. The study found that trained police officers hit their desired target less than 30% of the time [24]. We further assumed that civilian adults hit their desired target 10.0% less often than police officers[25]. Together, this implies that

$$\alpha_{ST} = c(0.3p_{police} + 0.2(1 - p_{police})) = 6.1\ corpses\ per\ day,$$

Accounting for the fact that not all contacts will be with zombies, we have that the ST intervention reduces the zombie population at the rate of

$$\alpha_{ST}\frac{Z}{H}.$$

Finally, because the ST intervention does not involve close-quarter combat, we assume $\beta_{ST} = 0$.

**2.3.3 The double-tap intervention.** Our DT intervention assumes adults use twice the amount of ammunition to kill zombies to ensure that they are dead. Thus, given two shots with a probability of 0.3 of hitting the target, the odds that a police officer successfully hits their target at least once is $1 - 0.7^2 = 0.51$. Similarly, for civilian adults, the probability of hitting their target at least once given two shots is $1 - 0.8^2 = 0.36$. It follows that

$$\alpha_{DT} = c(0.51p_{police} + 0.36(1 - p_{police})) = 10.97\ corpses\ per\ day,$$

Therefore, the DT intervention reduces the zombie population at the rate

$$\alpha_{DT}\frac{Z}{H}.$$

As the DT intervention does not involve close-quarter combat, we also assume $\beta_{DT} = 0$.

**2.4 Health benefit and cost-effectiveness analysis.** To determine the health benefits and costs of each zombie intervention relative to the baseline, we examined which intervention would be 1) the most beneficial in eradicating the zombie population by changing the stability of our models, 2) prevent most humans from becoming zombies, as measured by zombies averted, and 3) avert the most zombies and death, as measured by DALYs (due to zombieism) averted over two years. In addition, we also determine the most cost-effective intervention, as characterized by the ICER.

We estimate the additional effect of implementing weapons on the reproductive number of zombieism by

$$F = \beta + \frac{\beta_{weapon}}{H}A^*$$
$$V = \alpha_{baseline} + \frac{\alpha_{weapon}}{H}A^*$$

and

$$R_{eff} = FV^{-1} = \frac{\beta + \frac{\beta_{weapon}}{H}A^*}{\alpha_{baseline} + \frac{\alpha_{weapon}}{H}A^*}.$$

To quantify death and disability caused by zombies, we assume zombieism would have a similar level of disability as neurological sequelae. Therefore, we take the disability weight of zombieism, $D_z$, as 0.471 [22]. We calculated time discounted DALYS by combining the years lived with disability (YLD)

$$YLD = D_z \int_0^{730} (\frac{\beta}{H}Z(C + A + E) + \frac{\beta}{H}Z(C + A + E))e^{-rt} dt$$

with years of life lost

$$YLL = \int_0^{730} (\alpha_{baseline}Z + \frac{\alpha_{weapon}}{H}ZA)e^{-rt} dt.$$

It follows that the

$$DALYs = YLD + YLL.$$

Given the DALYs associated with each scenario, we measure the benefits relative to costs by

$$ICER = \frac{Cost_{intervention} - Cost_{baseline}}{DALYs_{intervention} - DALYs_{baseline}},$$

where $Cost_{intervention}$ is the price per capita associated with each intervention, and $Cost_{baseline}$ is the cost before an intervention, which we assume is 0 dollars. In addition, we assess each intervention to be cost-effective or very cost-effective using the criteria set forth by the World Health Organization [26], which stipulates that an ICER less than the per capita GDP is very cost-effective, and an ICER less than three times the per capita GDP is cost-effective.

**2.5 Intervention costs.** To coin our interventions as cost-effective, we estimated the cost associated with each of our interventions.

The ax intervention arms the adult population with steelhead axes, ranging in price from $30-55 and are widely available [27]. Taking the average ax price of $42.50, excluding taxes and distribution costs, an adult population of 227.3 million people gives a total intervention cost of 9.66 billion dollars.

For the ST intervention, the average 9mm bullets cost $0.40 per round, and one 9mm handgun costs on average $325 [28]. Based on human-to-human contact rates, we assume each adult comes in contact with at most 30 zombies per day, which leads to ammunition costs for the ST intervention of 2.01 trillion dollars. Thus, the price to arm civilian adults with 9mm handguns and ammunition is 2.09 trillion dollars, excluding taxes and distribution costs.

For the DT intervention, we assume that civilian adults use twice the amount of bullets as in the ST intervention, in addition to receiving a 9mm handgun. Thus, ammunition costs for the DT intervention are 4.02 trillion dollars, leading to a total intervention cost of 4.10 trillion dollars, excluding taxes and distribution costs.

**Results.**

Using our mathematical model of the zombie apocalypse, we evaluated three interventions aimed at preserving the human race and effectively killing off zombies. Specifically, the interventions comprised the use of steelhead axes, the ST intervention with a 9mm handgun, and the DT intervention with a 9mm handgun throughout two years. For each intervention, we also explored the effects of zombie speed, using slow, moderate, and fast transmission rates, and quantified health benefits and cost-effectiveness by the number of zombies averted, DALYs averted, and the ICER.

We determined conditions for the zombie apocalypse to die out based on $R_{eff}$ reaching the critical threshold of 1. Specifically, we examined how $R_{eff}$ changes with the effect of each intervention, namely $\alpha_{weapon}$. For ST and DT, we found that if $\alpha_{weapon} > 0.57$, $\alpha_{weapon} > 1.04$, and $\alpha_{weapon} > 1.32$ for slow, moderate speed, and fast zombies respectively, then zombies die out. For the ax intervention, slightly large values were required for zombies to die out, specifically $\alpha_{ax} > 0.58$, $\alpha_{ax} > 1.05$, and $\alpha_{ax} > 1.34$ for slow, moderate speed, and fast zombies, respectively.

At the baseline, our model predicted an annual 993.7, 996.7, and 997.2 DALYs per 1000 people with slow, moderate speed, and fast zombies, respectively. At the end of the two years, the baseline with zombies moving at a slow speed predicts 305 thousand humans remaining, moderate speed predicts 2040 humans remaining, and fast speed estimates 99 humans remaining.

Following the administration of steelhead axes to every adult human, our model predicted an annual DALYs averted of -1.403, -2.29, and -1.51 per 1000 people with slow, moderate speed, and fast zombies, respectively (Figure 1). In addition, 530 thousand humans remain with slow zombies, 3478 humans remaining with moderate-speed zombies, and 110 humans remaining with fast zombies (Table 2). Relative to the baseline, the steelhead ax intervention averts 225 thousand slow zombies, 1438 moderate speed zombies, and 12 fast zombies (Figure 2).

The ST intervention predicted an annual number of 993, 960, and 61 DALYs averted for low, moderate speed, and fast zombies, respectively (Figure 1). The model predicts 331.9 million humans remaining with slow zombies, 319.8 million humans remaining with moderate speed zombies, and 21 million humans remaining with fast zombies (Table 2). Relative to the baseline, the ST intervention averts 331.6 million, 319.8 million, and 21 million, slow, moderate speed, and fast zombies, respectively (Figure 2).

Our findings show the DT intervention would cause an annual number of 993, 965, and 193 DALYs averted per 1000 people with slow, moderate speed, and fast zombies, respectively (Figure 1). In addition, given this intervention, we predict 331.9 million humans remain with slow zombies, 321.4 million humans for moderate speed zombies, and 64.8 million humans with fast zombies (Table 2). Given these numbers, the DT intervention would avert 331.6 million, 321.4 million, and 64.8 million slow, moderate speed, and fast zombies, respectively (Figure 2).

We found the ax intervention scenario to be ineffective. A significant number of zombies are averted in this scenario, however, at the expense of human lives. To elaborate, the ICER values are -30.3 thousand, -18.5 thousand, and -28.2 thousand with slow, moderate speed, and fast zombies, respectively (Table 2).

Upon comparing the ST and DT interventions' ICER values to the United States GDP per capita of 65,543.6, we found both the ST intervention to be very cost-effective when the zombies are moving at slow and moderate speeds, and not cost-effective when zombies are moving at fast speeds. Similarly, we found DT to be very cost-effective when zombies move at slow and moderate speeds, but only cost-effective when zombies move faster. To elaborate, the ICER values for these interventions at slow, moderate, and fast zombie speeds are 4.79 thousand, 4.96 thousand, and 78.1 thousand for ST, and 9.25 thousand, 9.53 thousand, and 47.7 thousand for DT. Despite DT killing a slightly higher amount of zombies than ST, ST is more favorable because it kills nearly the same amount of zombies as DT and costs half the price.

**Table 2. Zombies, Zombies averted, DALYs averted, and ICER.**

|  | Slow | Moderate speed | Fast |
|---|---|---|---|
| Baseline |  |  |  |
| Total zombies | $3.32 \cdot 10^8$ | $3.33 \cdot 10^8$ | $3.33 \cdot 10^8$ |
| Total DALYs | 993.7 | 996.7 | 997.2 |
| Total deaths | $3.32 \cdot 10^8$ | $3.33 \cdot 10^8$ | $3.33 \cdot 10^8$ |
| Steelhead ax intervention |  |  |  |
| Total zombies | $3.32 \cdot 10^8$ | $3.33 \cdot 10^8$ | $3.33 \cdot 10^8$ |
| Zombies averted per year | $2.32 \cdot 10^5$ | $2.20 \cdot 10^3$ | 92 |
| Annual DALYs averted per 1000 people | $-1.40$ | $-2.29$ | $-1.51$ |
| Deaths averted per year | $2.26 \cdot 10^5$ | $1.45 \cdot 10^3$ | 11 |
| ICER | $-3.03 \cdot 10$ | $-1.85 \cdot 10^4$ | $-2.82 \cdot 10^4$ |
| ST intervention |  |  |  |

| | Slow | Medium | Fast |
|---|---|---|---|
| Total zombies | $8.18 \cdot 10^3$ | $1.22 \cdot 10^7$ | $1.97 \cdot 10^8$ |
| Zombies averted per year | $3.32 \cdot 10^8$ | $3.21 \cdot 10^8$ | $2.10 \cdot 10^7$ |
| Annual DALYs averted per 1000 people | 993 | 960 | 61 |
| Deaths averted per year | $3.32 \cdot 10^8$ | $3.20 \cdot 10^8$ | $2.10 \cdot 10^7$ |
| ICER | $4.79 \cdot 10^3$ | $4.96 \cdot 10^3$ | $7.81 \cdot 10^4$ |
| DT intervention | | | |
| Total zombies | $7.61 \cdot 10^3$ | $1.05 \cdot 10^7$ | $2.68 \cdot 10^8$ |
| Zombies averted per year | $3.32 \cdot 10^8$ | $3.22 \cdot 10^8$ | $6.49 \cdot 10^7$ |
| Annual DALYs averted per 1000 people | 993 | 965 | 193 |
| Deaths averted per year | $3.32 \cdot 10^8$ | $3.22 \cdot 10^8$ | $6.49 \cdot 10^7$ |
| ICER | $9.25 \cdot 10^3$ | $9.53 \cdot 10^3$ | $4.77 \cdot 10^4$ |

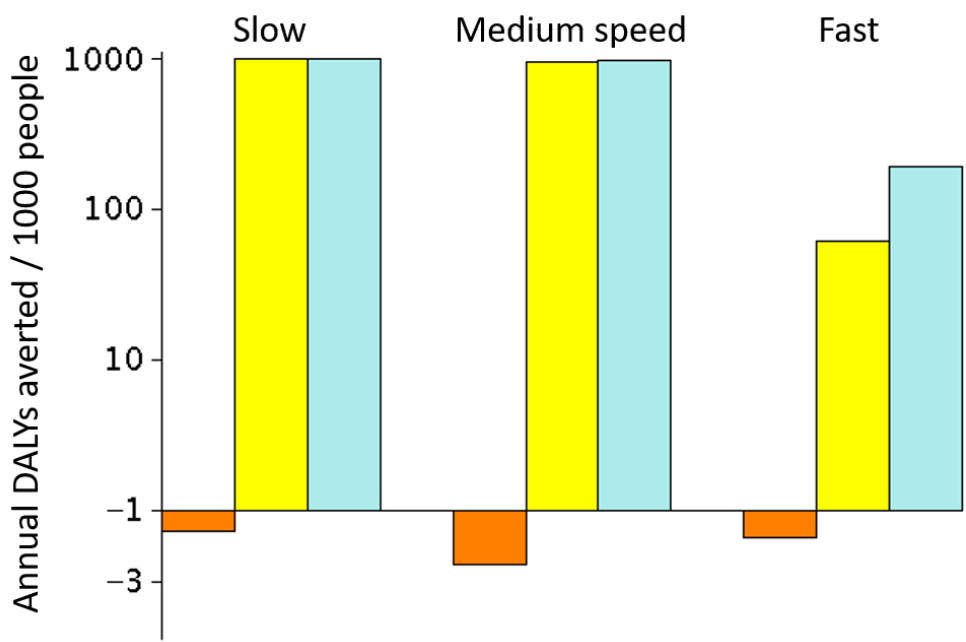

**Figure 1.** Annual zombieism DALYs averted. The number of DALYs averted for ax (orange), ST (yellow) and DT (blue) interventions. Forecasted the number of DALYs averted for each intervention at every zombie speed.

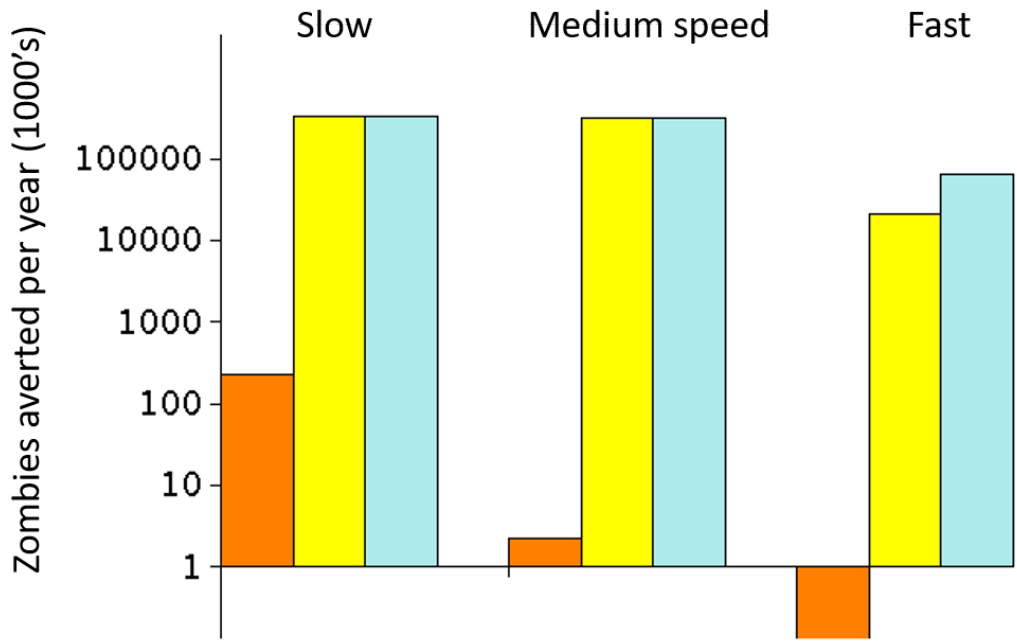

**Figure 2.** Zombies averted per year. The number of zombies averted for the ax (orange), ST (yellow) and DT (blue) interventions. Forecasted the number of zombies averted for each intervention at every zombie speed.

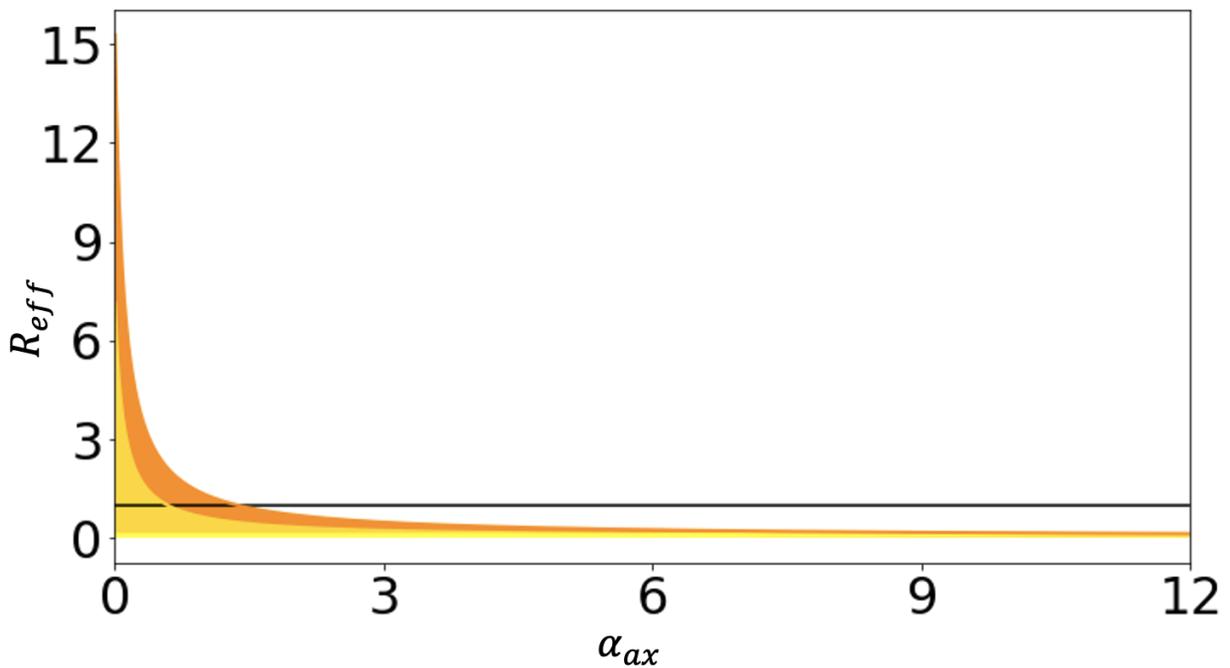

**Figure 3.** Effective reproductive number versus alpha weapon. Ax intervention with slow (gold), moderate speed (yellow), and fast zombies (dark orange).

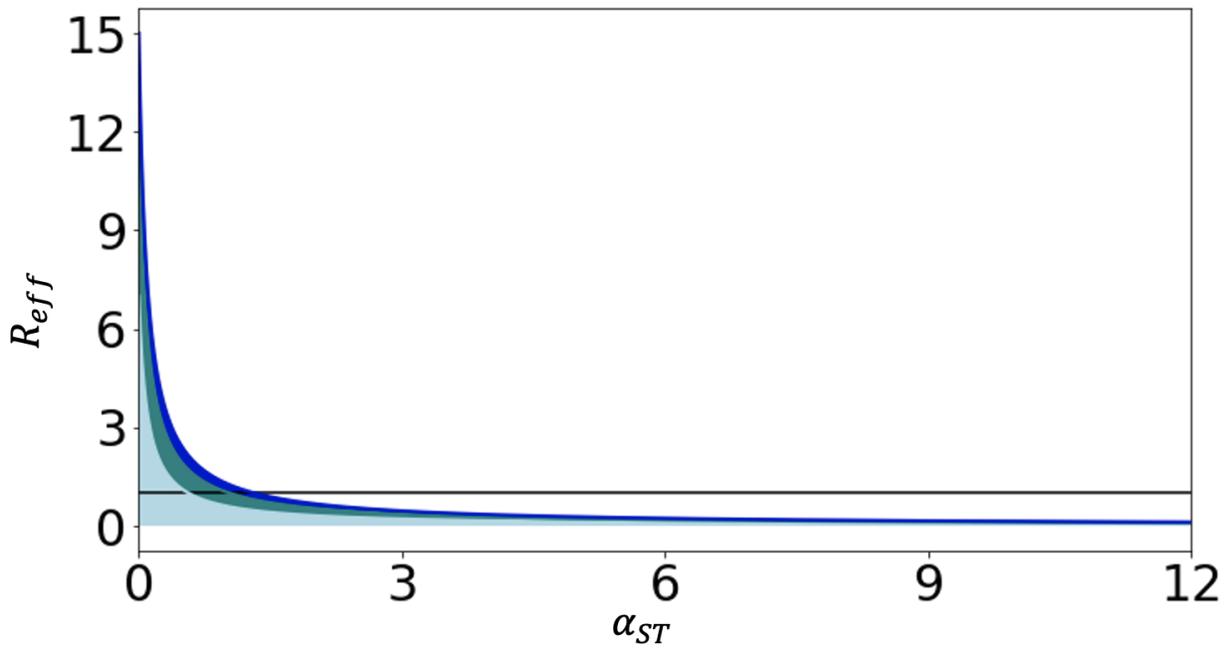

**Figure 4.** Effective reproductive number versus alpha weapon. ST intervention with slow (light blue), moderate speed (teal), and fast zombies (dark blue).

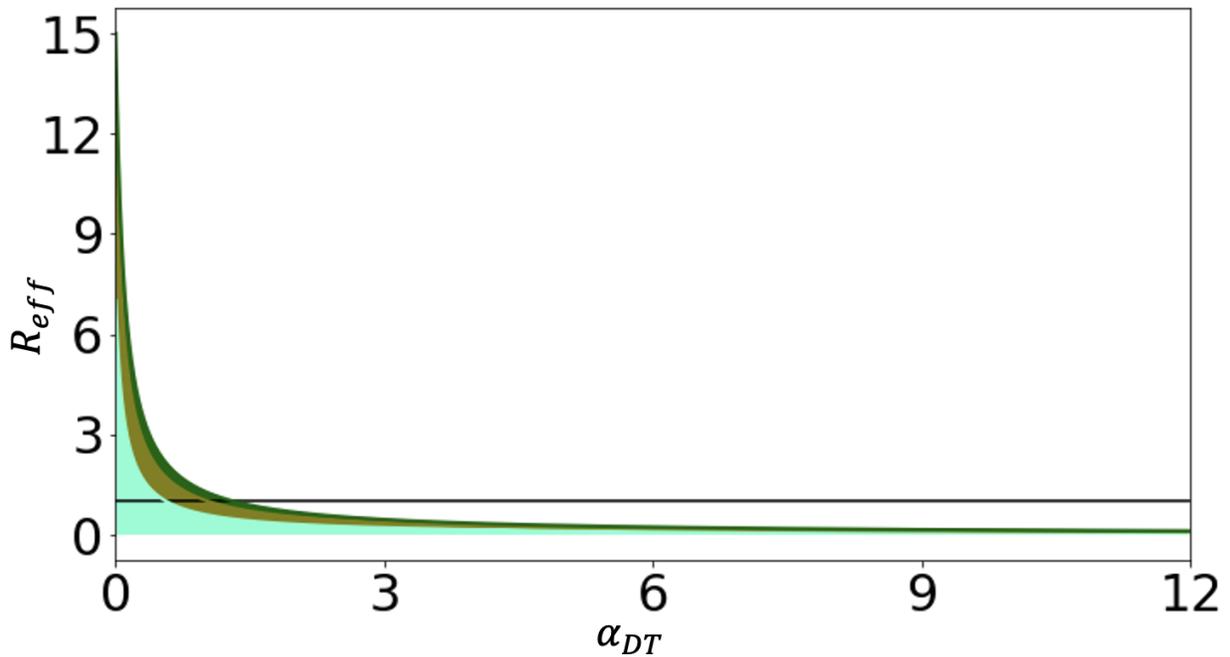

**Figure 5.** Effective reproductive number versus alpha weapon. DT intervention with slow (aquamarine), moderate speed (olive), and fast zombies (dark green).

**Discussion.**

Using our mathematical model, we evaluated interventions for fighting zombie outbreaks in the United States. We found the ST intervention to be the most cost-effective intervention as it kills nearly the same number of zombies as the DT intervention at almost half the price. We also found that arming the entire adult population with steelhead axes is counterproductive, as it lowers the number of zombies at the expense of human lives. Regarding the health burden of zombies, the ST and DT interventions provided similar amounts of DALYs zombies, and deaths averted over two years.

Although extremely cheap, the steelhead ax intervention killed zombies at the expense of human lives and cost DALYs instead of saving them. While at first, the negative amount of DALYs averted may seem surprising, the increased risks posed by melee combat with zombies seem plausible for causing this expenditure. Thus, we advise caution with any form of intervention that promotes close-quarter warfare, especially against fast-moving zombies, and suggest such a means only be used as a last resort for survival.

We discovered that the majority of our interventions using 9mm handguns were either cost-effective or very cost-effective in the United States. Regarding the ST intervention, we found it was very cost-effective while zombies were moving at slow and moderate speeds, yielding ICER values of 4.79 and 4.96 thousand, respectively. While these values are low compared to the American GDP per capita, they may not be cost-effective in many underdeveloped countries that feature substantially lower GDPs. Particularly in under-developed countries in Sub-Saharan Africa, where the GDP is a fraction of the American GDP. In comparison, the interventions are likely cost-effective or very cost-effective in developed countries, such as Canada and select European countries, due to GDPs similar to the US.

Many mathematical models have been created to predict zombie outbreaks [10,29–31]. As such, our baseline simulations agree with the majority of these models, in that they indicate zombies overpowering humankind in a short timeframe. In contrast, simulations of our ST and DT interventions illustrate a more positive outlook for when humans fight back against zombies, in that the zombie population dies out, rather than causing an endless cycle [10]. While either situation is plausible, advances in medicine, public health infrastructure, and military sciences provide optimism for humankind's ability to curtail any such outbreak.

We showed how ax and gun interventions impact zombie apocalypses in the US. However, with modest modifications, our work could provide insight into additional regions where the zombie outbreak may occur, as it is not likely a zombie apocalypse would be restricted to the United States. Towards this regard, including geographical aspects of transmission, both within the US and externally, would stand to more accurately predict the spread of zombies across the country and the world, in addition to improving projects on the effectiveness and affordability of our interventions. Similarly, another avenue for investigation is forecasting the effects of other interventions, such as quarantine or vaccination, or a combination of interventions, as such strategies typically prove more practical and cost-effective than any single intervention alone. In addition to alternative zombie interventions, a potentially interesting of our model would be to include zoonotic transmission. To elaborate, in the films Resident Evil and I am Legend [32–34], animals can become zombies. Thus, as zoonotic transmission depends on vector-host interactions, the associated disease dynamics would likely be very complex [35,36], making any public health efforts for controlling a zombie apocalypse much more complicated.

Inevitably, our model has limitations. At the forefront, human zombies do not exist, although they do for other species, such as ants [37]. Due in part to this limitation, we made assumptions on the transmission rate and basic reproductive number of zombieism, which were based upon historical data of highly infectious diseases. We also assumed adults were able to kill zombies upon contact with a steelhead ax and 9mm handgun, which would likely vary significantly in practice. Towards this regard, we do not fully account for the heterogeneity of population demographics and only consider the human population as children, adults, or elderly. In the United States today, approximately 26% of the entire adult population lives with either a physical or mental disability [38]. Because of this high number, it is not reasonable to assume all members of the adult population would have the same likelihood of killing a zombie.

If a zombie apocalypse occurs in the United States, the government would be tasked with protecting the population and ensuring humanity's survival with limited resources. Based on our analysis of interventions, the ST would be the best option for ammo conservation, affordability, and effectiveness. This intervention would prevent millions of human deaths, is very cost-effective for slow and moderate speed zombies, and costs about half of the DT intervention while yielding almost identical results. Because of these advantages, we suggest that governments warding off zombie apocalypses should equip citizens with ranged weapons, while encouraging them to not be wasteful in the use of ammunition and to avoid close-quarter combat.

## Availability of data and materials

All data generated or analyzed during this study are included in the is published article and its supplementary information files

## Competing interests

The authors declare that they have no competing interests.

## Funding

This work was supported by the Mathematical Association of America National Research Experiences for Undergraduates Program grant 897.

## Author's contributions

JP, and AR, co-wrote the first draft of the manuscript. JP, AR, KT and SG revised and edited subsequent versions. JP conceptualized the mathematical model and performed the literature review that determined model parameters and facilitated the discussion section of the manuscript. AR implemented the mathematical model in software, and conducted the cost-effectiveness analysis. SG conceptualized the premise of the work. All authors read and approved the final manuscript.

**Acknowledgements**

Not applicable

**Student Biographies**

Ahmani Roman: Ahmani Roman is currently enrolled in the applied physics program at Siena College. He intends to continue his studies in the M.S. Mechanical Engineering program at Clarkson Graduate School.

Jacob Pacheco: Jacob Pacheco graduated from Siena College in 2021 with a B.A. in Mathematics. He is currently enrolled in the Graduate Data Analytics program at Siena College.